# Giant spin-to-charge conversion in germanium tin epilayers

S. Oyarzún,[1] C. Gonzalez-Fuentes,[2] Erick Burgos[1], M. Myronov,[3] M. Jamet,[4] R. L. Rodríguez-Suárez,[2] and F. Pezzoli[5].

[1] *Departamento de Física, CEDENNA, Universidad de Santiago de Chile (USACH),Chile*

[2] *Facultad de Física, Pontificia Universidad Católica de Chile, Avenida Vicuña Mackenna 4860, Casilla 306, Santiago, Chile.*

[3] *Department of Physics, The University of Warwick, Coventry CV4 7AL, United Kingdom*

[4] *Univ. Grenoble Alpes, CEA, CNRS, IRIG-Spintec, F-38000 Grenoble, France*

[5] *Dipartimento di Scienza dei Materiali, Università degli Studi di Milano-Bicocca and BiQuTe, via Cozzi 55, I-20125 Milano, Italy*

We report a study of the spin-to-charge current conversion in compressively strained $Ge_{1-x}Sn_x$ alloy epilayers as a function of the Sn concentration by means of the inverse spin Hall effect (ISHE). The spin current is generated by spin-pumping effect (SPE) from a thin NiFe layer driven into ferromagnetic resonance (FMR). By simultaneously measuring the magnetic damping of the NiFe layer and the ISHE-induced charge current ($I_{ISHE}$) we extract two key spintronics parameters: the spin Hall angle $\theta_{SH}^{GeSn}$ and the effective spin mixing conductance. Our results reveal a giant spin-to-charge conversion and a non-monotonic dependence of the $I_{ISHE}$ signal on the Sn concentration, consistent with the variation in the magnetic damping observed in FMR. The values of $\theta_{SH}^{GeSn}$ are comparable to those reported for heavy metals such as Pt and Ta. Furthermore, we show that the spin conductivity at the Au/GeSn interface can be enhanced by tuning the Sn concentration.

# I. INTRODUCTION

One major challenge in semiconductor spintronics devices is achieving three key functions: generating, manipulating, and detecting spin currents [1]. An effective method of generating a spin current is through the spin pumping effect (SPE) [2,3]. In this phenomenon, when a ferromagnet (FM) is driven into ferromagnetic resonance (FMR), its precessing magnetization injects a spin current into an adjacent non-magnetic (NM) or semiconductor (SC) layer. The spin current generated by SPE can be detected by two main manifestations: (i) its conversion into a charge current in the NM or SC layer via the inverse spin Hall effect (ISHE) [4], and (ii) a change of the magnetic damping of the FM layer [5], resulting from the flow of the angular momentum out of the ferromagnet. The efficiency of the spin-to-charge interconversion is particularly important in semiconductors, as it enables the integration of spintronics functionalities with conventional electronic technologies. The incorporation of semiconductors such as silicon or germanium in the realization of spin-based transistors represents a cornerstone in the development of next-generation microelectronic devices [6]. Among these materials, Ge exhibits higher carrier mobility and stronger spin-orbit coupling compared to Si, while remaining compatible with mainstream microelectronics.

The physical parameter that quantifies spin-to-charge conversion is the spin Hall angle $\theta_{SH}$, which can be experimentally determined through SPE and ISHE measurements [7]. To date, various efforts have been made to explore the spin-to-charge conversion in Ge [8–11]; however, its spin Hall angle remains significantly smaller than that of heavy metals such as Pt or Ta. Recently, epitaxially grown germanium-tin ($Ge_{1-x}Sn_x$) binary alloys have emerged as a promising semiconductor for spintronics applications [12–14]. Several research groups have investigated the injection of spin currents and their conversion into charge currents via ISHE, leveraging the tunable spin–orbit interaction as a function of Sn content [15–18].

In this work, we report a very efficient spin-to-charge conversion by means of the ISHE in Au/NiFe/Au/$Ge_{1-x}Sn_x$/Ge/Si heterostructures as a function of the Sn concertation. The spin current was generated in the NiFe layer through FMR, injected into the $Ge_{1-x}Sn_x$ layer, and converted into charge current, which was measured as a drop voltage in open circuit conditions. The simultaneous measurements of the magnetic damping of the NiFe layer and the generated charge current by the ISHE allow the quantitative estimation of the spin Hall angle ($\theta_{SH}^{GeSn}$) and the effective spin mixing conductance. The magnitude obtained for $\theta_{SH}^{GeSn}$ is comparable to those found in NiFe/heavy-metal interfaces and nearly two orders of magnitude higher than that of Ge.

## II. EXPERIMENTAL DETAILS

For this experiment, a set of samples of Ge$_{1-x}$Sn$_x$/Ge/Si heterostructures were grown using a reduced pressure chemical vapour deposition (RP-CVD) system. All Ge$_{1-x}$Sn$_x$ epilayers were grown to be intentionally compressively strained with Sn content, and the thickness varied from as low as 3 % up to 13% and from 24 nm up to 63 nm, respectively. The Ge$_{1-x}$Sn$_x$ epilayers were grown at ~0.1 nm/s growth rate on 150 mm diameter Si(001) wafers via a non-intentionally doped relaxed Ge buffer layer. All the heterostructures were characterized by XTEM images and HR-XRD analytical techniques to obtain epilayers thicknesses, degree of strain in them and Sn content in the Ge$_{1-x}$Sn$_x$ epilayers. More details about the epitaxial growth and structural characterizations of such films can be found in Ref. [16,17,19]. NiFe/Au thin films were subsequently deposited by e-beam evaporation onto the Ge$_{1-x}$Sn$_x$/Ge/Si heterostructures to perform spin pumping and inverse spin Hall effect (ISHE) experiments.

### a) SPIN PUMPING AND INVERSE SPIN HALL EFFECT EXPERIMENT

To study the spin-to-charge current conversion in the NiFe/Au/Ge$_{1-x}$Sn$_x$/Ge/Si multilayers, we measured the spin-pumping-ISHE-voltage under FMR conditions. Figure 1(a) shows the FMR-driven spin pumping process with the magnetization of the NiFe layer exited by a rf magnetic field $h_{rf}$ at a frequency of 9.68 GHz under a static magnetic field $\mu_0 H$. When the FMR condition is fulfilled at $H = H_R$, the precessing magnetization generates a spin accumulation at the NiFe/Au interface that diffuses injecting a net spin current into the adjacent Au/Ge$_x$Sn$_{1-x}$ layers. The spin current $\boldsymbol{j}_S = j_S \hat{\boldsymbol{z}}$ polarized along the $\hat{\boldsymbol{x}}$ direction (the magnetization-precession axis, see also Fig.1 a) generates a transverse electric field $\boldsymbol{E}_{ISHE} \propto (\boldsymbol{j}_S \times \hat{\boldsymbol{x}})$ in the structure due to the ISHE [7] that, in turn, is converted into a dc voltage $V_{ISHE} \propto |\boldsymbol{j}_S \times \hat{\boldsymbol{x}}|_y$ at the sample edges where $|\boldsymbol{j}_S \times \hat{\boldsymbol{x}}|_y$ represents the $y$ component of $\boldsymbol{E}_{ISHE}$. The $V_{ISHE}$ voltage is measured directly with a nanovoltmeter connected by coaxial wires to electrodes attached at the edges of the sample, as shown in Fig. 1(a).

Figures 1(b) and 1(c) show, respectively, the derivative of the FMR absorption spectrum for the NiFe(15nm)/Au(5nm)/Ge$_{0.87}$Sn$_{0.13}$(40nm) sample measured at a microwave power of 200 $mW$ and the simultaneous measurement of the ISHE induced voltage $V_{ISHE}$ vs. $\mu_0 H$ for several microwave power levels and two directions of the in-plane magnetic field (parallel and antiparallel to the $x$-direction). As expected, and according to the relation $V_{ISHE} \propto |j_S \times \hat{x}|_y$, the sign of the voltage changes with the field reversal.

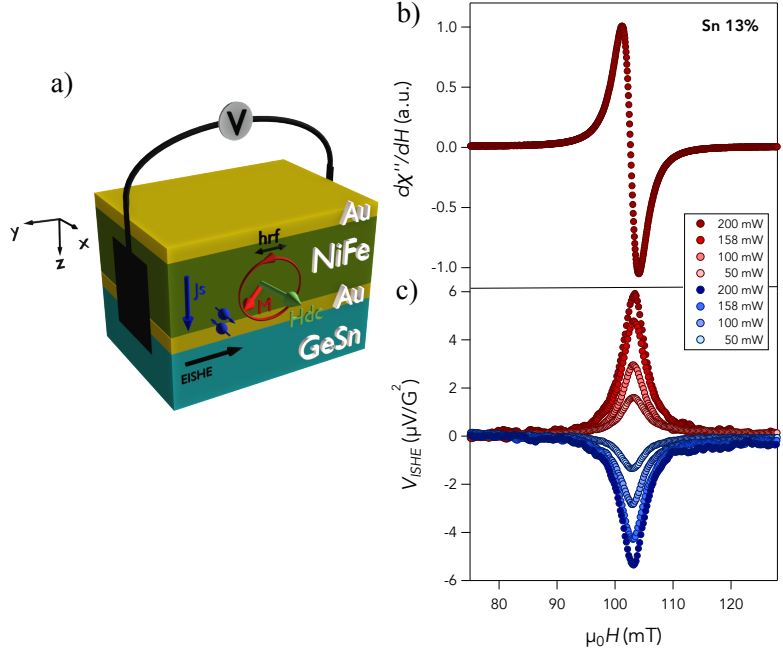

Fig. 1. a) Sketch of the NiFe/Au/Ge$_x$Sn$_{1-x}$ three-layer and experimental arrangement for spin pumping-transverse voltage measurements. b) FMR absorption derivative versus the magnetic field, measured at 9.68 GHz and microwave power of 200 mW. c) ISHE-induced voltages measured at several microwave power levels for two directions of the in-plane magnetic field (parallel and antiparallel to the x-direction).

Under FMR conditions, the induced voltage as a function of the applied magnetic field can be written as a combination of symmetric and asymmetric components [20,21]

$$V = V_S \frac{\Delta H^2}{(H - H_R)^2 + \Delta H^2} + V_A \frac{\Delta H (H - H_R)}{(H - H_R)^2 + \Delta H^2} \qquad (1)$$

being, $V_S$ and $V_A$ the respective amplitudes, $\mu_0 H_R$ the resonance field and $\mu_0 \Delta H_{pp}$ the FMR peak-to-peak linewidth. In ferromagnetic (FM) layers, both contributions have their origin in the anisotropic magnetoresistance (AMR) effect [20–23]. Meanwhile, in ferromagnetic/non-magnetic (FM/NM) bilayers, the voltage induced by the spin pumping effect has a symmetric contribution, and in Eq.(1) the corresponding contribution is $V_S = V_{ISHE} + V_{S-AMR}$. Both, symmetric and asymmetric contributions are proportional to the microwave power ($\propto h_{rf}^2$) and can be extracted from the fit of the dc voltage signals.

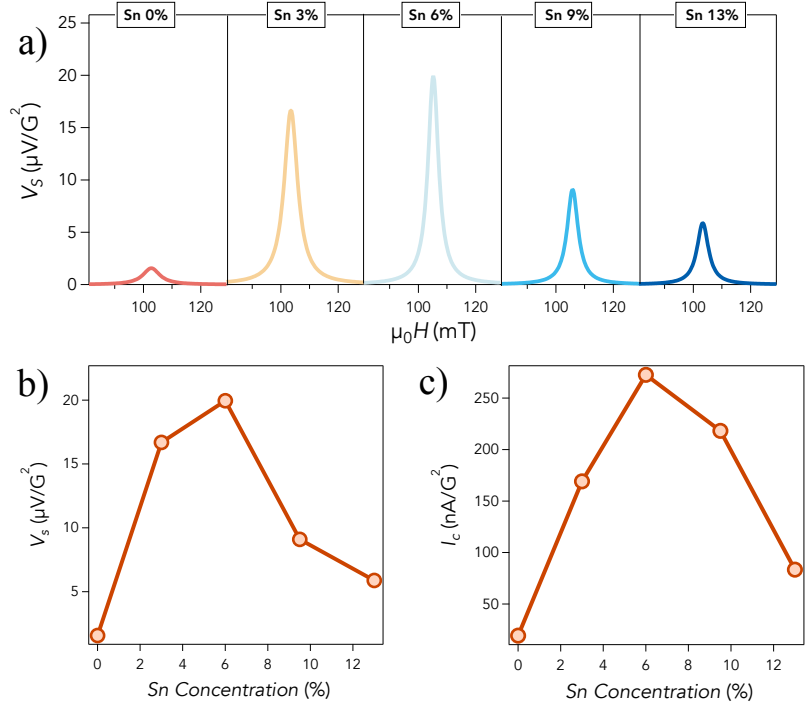

Fig. 2. Amplitude of the symmetric part of the induced voltage versus magnetic field H for different concentrations of Sn (x=0, 3, 6, 9 and 13 %) in NiFe(15nm)/Au(5nm)/Ge$_{1-x}$Sn$_x$(40nm). (b), and (c) show the amplitude $V_S$, and the corresponding current versus the Sn concentration.

Figure 2(a) shows the field scan of the symmetric signal $V_S$ for the set of five samples Py(15 $nm$)/Au(5 $nm$)/Ge$_{1-x}$Sn$_x$ for different concentrations of Sn ($x = 0, 3, 6, 9$ and 13 %). The amplitude of $V_S$ as a function of the Sn concentration is shown in Fig. 2(b). In the entire set of samples $V_{A-AMR} \ll V_S$ and the electrical resistance is dominated by the NiFe/Au part of the stack; we attribute the symmetric voltage signal to the ISHE. Additionally, as $V_S$ depends on the electrical resistance of the structure, the ISHE current $I_C$ ($= V_S/R$) shown in Fig. 2(c) is the physically most relevant quantity. As observed, there is a non-monotonous dependence of both $V_S$ and $I_C$ on the Sn concentration with maximum amplitude values of 19.8 ($\mu V/G^2$) and 272 ($nA/G^2$) at Sn=6%, respectively. The most remarkable result in Fig. 2(c) is the nearly tenfold increase in $I_{C(Sn=6\%)}/I_{C(Sn=0)}$, representing a significant advancement in the search for methods to enhance the spin-charge interconversion in materials compatible with mainstream silicon technology.

We investigated the dynamic magnetic response of NiFe/Au/Ge₁₋ₓSnₓ/Ge/Si multilayers using a broadband ferromagnetic resonance (FMR) setup. The samples were placed on a coplanar waveguide, which excited magnetization dynamics in the NiFe layer within the 4–14 GHz frequency range under an in-plane magnetic field configuration. The resonance field $H_R$ and the peak-to-peak linewidth $\Delta H_{pp}$ were extracted by fitting the derivative of the microwave absorption spectra with a Lorentzian function at each frequency ($f$). We determined the effective magnetization $M_{eff}$ from the frequency dependence of the resonance field (Fig. 3 (b)) by applying the well-known Kittel`s equation, while the linear variation of $\Delta H_{pp}$ as a function of frequency (Fig. 3 (a)) provided insight into the Gilbert damping parameter $\alpha$ and extrinsic broadening mechanisms $\Delta H_{\text{inh}}$, according to

$$\Delta H_{pp} = \frac{2}{\sqrt{3}}\left(\frac{2\pi f}{\gamma}\right)\alpha + \Delta H_{\text{inh}} \tag{2}$$

with $\gamma = \frac{g_{eff}\mu_B}{\hbar}$, $\hbar$ is the reduced Planck constant, $g_{eff}$ the Landé factor and $\mu_B$ the Bohr magneton.

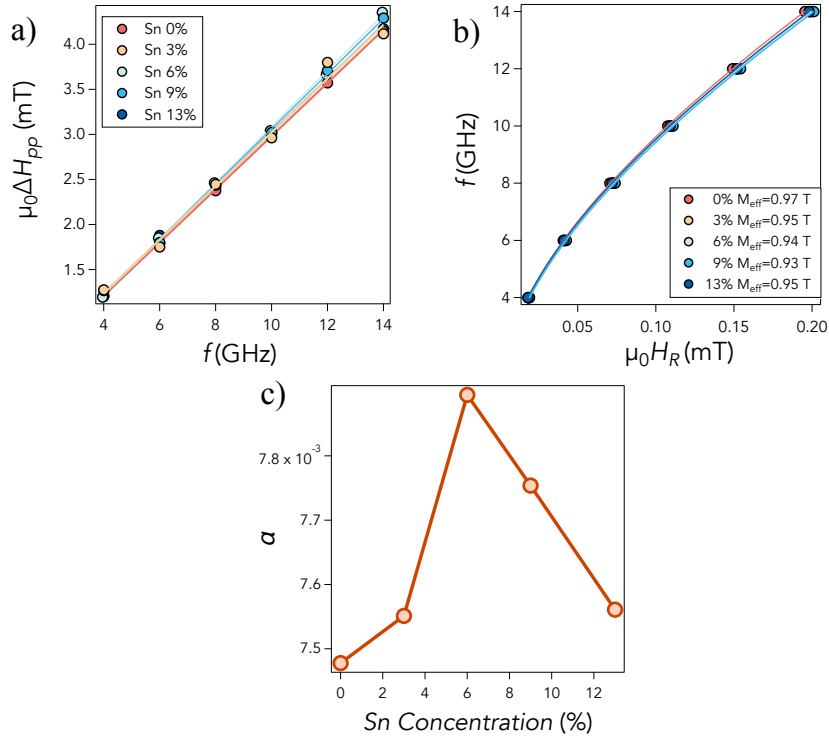

Fig. 3. a) Peak-to-peak FMR linewidth $\mu_0\Delta H_{pp}$ as a function of the excitation frequency, dots are experimental data, and the solid line is a fit using Eq. (2). b) Excitation frequency as a function of the resonance field $\mu_0 H_R$, dots are experimental data, and the solid line is a fit using the well-know Kittel equation. c) Damping parameter as a function of Sn concentration obtained from a fit using Eq. (2).

In Fig. 3(c), we observe a non-monotonous behavior of the Gilbert damping parameter as a function of Sn concentration. Furthermore, the damping parameter reaches a maximum at a Sn concentration of 6%. Both features are consistent with the trends observed in the spin-to-charge conversion measurements. The effective magnetization exhibits similar values across the range of Sn concentrations.

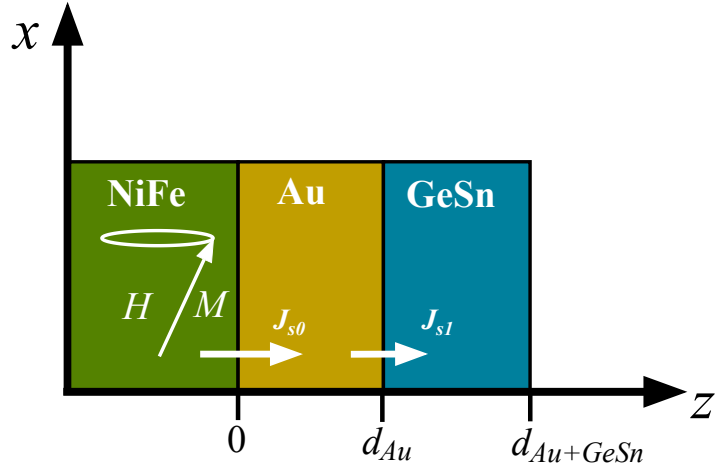

Fig. 4. Scheme of the NiFe/Au/GeSn/Ge multilayer and FMR spin pumping into the adjacent normal metal Au layer. The total spin current at the NiFe/Au interface diffuses through the other layers.

To understand the data in Fig. 2 it is necessary to consider the spin-pumping diffusion and spin-to-charge current conversion in the NiFe/Au/GeSn multilayer (see Fig. 4). Under radiofrequency excitation, the precessing magnetization of the NiFe layer pumps spins through the NiFe/Au interface ($y = 0$), generating a spin current density given by [24,25]

$$j_{s0} = \frac{\hbar \omega p g_{eff}^{\uparrow\downarrow}}{4\pi}\left(\frac{h_{rf}}{\Delta H}\right)^2 L(H - H_R), \qquad (3)$$

where $\hbar$ is the reduced Planck constant, and $g_{eff}^{\uparrow\downarrow}$ is the real part of the effective spin mixing conductance of the NiFe/Au/GeSn structure that considers both the emitted spin current and spin backflow. $\Delta H (= \alpha \omega / \gamma)$ and $H_R$ are, respectively, the linewidth and FMR field of the NiFe layer at the frequency $\omega$, $L(H - H_R) = L(H - H_R) = \Delta H^2 / [(H - H_R)^2 + \Delta H^2]$ is the normalized Lorentzian line shape, and $p$ is the precession ellipticity [26].

The real part of the effective spin-mixing conductance in Eq.(3), can be written as

$$g_{eff}^{\uparrow\downarrow} = \frac{g_r^{\uparrow\downarrow}}{1 + \beta g_r^{\uparrow\downarrow}}, \tag{4}$$

where $g_r^{\uparrow\downarrow}$ is the spin mixing conductance of the NiFe/Au interface, and $\beta$ is the spin backflow factor [27,28], which, for the structure shown in Fig. 4 is

$$\beta = \frac{1}{g_{Au}} \left[ \frac{\frac{g_{GeSn}}{g_{Au}} \tanh d_{GeSn}/\lambda_{GeSn} \tanh d_{Au}/\lambda_{Au} + 1}{\frac{g_{GeSn}}{g_{Au}} \tanh d_{GeSn}/\lambda_{GeSn} + \tanh d_{Au}/\lambda_{Au}} \right] \tag{5}$$

where $g_i = \sigma_i h/\lambda_i e^2$ is the conductance (per spin, in units of $m^{-2}$), and $\sigma_i$, $d_i$ and $\lambda_i$ are the electrical conductivity, thickness, and spin diffusion length of the $i$-layer, respectively.

The parameters defining the spin diffusion process and the spin-to-charge current conversion are respectively, (i) the effective spin mixing conductance $g_{eff}^{\uparrow\downarrow}$ at the NiFe/Au interface and (ii) the spin Hall angle, which measures the strength of the ISHE. The effective spin mixing conductance $g_{eff}^{\uparrow\downarrow}$ in Eq.(4) can be determined from the measurement of the Gilbert damping parameter for the sample in which Sn = 0 %. With, $\lambda_{Au} = 32\ nm$ (see below), $d_{Au}/\lambda_{Au} \ll 1$, and from Eq. (5) $g_{eff}^{\uparrow\downarrow}(S_n = 0\%)$ reduces to

$$g_{eff}^{\uparrow\downarrow}(S_n = 0\%) = \frac{g_r^{\uparrow\downarrow}}{1 + \frac{g_r^{\uparrow\downarrow}}{g_{Ge}} \left( \frac{g_{Ge}}{g_{Au}} \frac{d_{Au}}{\lambda_{Au}} \tanh\left(\frac{d_{Ge}}{\lambda_{Ge}}\right) + 1 \right) \left( \tanh\left(\frac{d_{Ge}}{\lambda_{Ge}}\right) + \frac{g_{Au}}{g_{Ge}} \frac{d_{Au}}{\lambda_{Au}} \right)^{-1}} \tag{6}$$

In terms of materials parameters, $\lambda_{Ge} = 1000\ nm$, $\sigma_{Au} = 4.1 \times 10^7\ (\Omega\ m)^{-1}$, and $\sigma_{Ge} = 2.7 \times 10^4\ (\Omega\ m)^{-1}$ [9], $g_{Ge} d_{Au}/g_{Au}\lambda_{Au} = \sigma_{Ge} d_{Au}/\sigma_{Au}\lambda_{Ge} \approx 2.5 \times 10^{-6} \ll 1$, $g_{Au} d_{Au}/g_{Ge}\lambda_{Au} = \sigma_{Au}\lambda_{Ge} d_{Au}/\sigma_{Ge}\lambda_{Au}^2 \approx 6.1 \times 10^3 \gg 1$ and Eq. (6) yields

$$g_{eff}^{\uparrow\downarrow}(Sn = 0\%) = \frac{g_r^{\uparrow\downarrow}}{1 + \beta_{0\%} g_r^{\uparrow\downarrow}} \approx \frac{g_r^{\uparrow\downarrow}}{1 + \left(\frac{g_r^{\uparrow\downarrow}}{g_{Au}}\right)\left(\frac{\lambda_{Au}}{d_{Au}}\right)}. \tag{7}$$

The Gilbert damping is $\alpha_{Sn0\%} = \alpha_0 + \alpha'$, where $\alpha_0$ is the intrinsic contribution and $\alpha'$ is the additional damping due to spin pumping given by [28]

$$\alpha' = \frac{\gamma \hbar g_{eff}^{\uparrow\downarrow}(S_n = 0\%)}{4\pi M_{eff} t_{FM}}. \tag{8}$$

Here, $M_{eff}$ and $t_{FM}$ are the effective magnetization and thickness of the NiFe layer, respectively, and $\gamma$ is the gyromagnetic ratio. With the following material parameters: $\gamma = 17.6\ GHz/kOe$, $4\pi M_{eff} = 10.4\ kG$, $\alpha_0 = 7.0 \times 10^{-3}$, and $g_r^{\uparrow\downarrow} = 2 \times 10^{19}\ m^{-2}$ [29], with the experimental value of the Gilbert damping at zero Sn concentration (see Fig. (1d)), $\alpha_{Sn0\%} = 7.47 \times 10^{-3}$, from Eqs. (7) and (8) we obtain $\lambda_{Au} = 32\ nm$ and $g_{eff}^{\uparrow\downarrow}(Sn = 0\%) = 4.1 \times 10^{18}\ m^{-2}$.

Once $\lambda_{Au}$, and $g_{eff}^{\uparrow\downarrow}(Sn = 0\%)$ have been determined, we can obtain the spin-Hall angle $\theta_{SH}^{Au}$ of the Au layer. In the NiFe/Au/Ge system, the spin current injected into the Au/Ge layers decays along the $z$ direction due to the spin relaxation as

$$j_{sAu}(z) = j_{s0} \left( \frac{\frac{g_{Ge}}{g_{Au}} \tanh(d_{Ge}/\lambda_{Ge}) \cosh[(d_{Au} - z)/\lambda_{Au}] + \sinh[(d_{Au} - z)/\lambda_{Au}]}{\frac{g_{Ge}}{g_{Au}} \tanh(d_{Ge}/\lambda_{Ge}) \cosh(d_{Au}/\lambda_{Au}) + \sinh(d_{Au}/\lambda_{Au})} \right) \tag{9}$$

$$j_{sGe}(z) = j_{s0} \frac{g_{Ge}}{g_{Au}} \left( \frac{\sinh[(d_{Au} + d_{Ge} - z)/\lambda_{Ge}]}{\frac{g_{Ge}}{g_{Au}} \sinh(d_{Ge}/\lambda_{Ge}) \cosh(d_{Au}/\lambda_{Au}) + \cosh(d_{Ge}/\lambda_{Ge}) \sinh(d_{Au}/\lambda_{Au})} \right) \tag{10}$$

where the spin current density $j_{s0}$ at the interface $z = 0$ is given by Eq.(3).

The spin currents $j_{sAu}(z)$ and $j_{sGe}(z)$ in Eqs. (9) and (10) are converted by means of the ISHE into a charge current $I_c \equiv l[d_{Au}\langle j_c^{Au}\rangle + d_{Ge}\langle j_c^{Ge}\rangle]$ with $l = 0.4\ mm$ and $\langle j_c^{Au,Ge}\rangle$ are the length of the stack (see Fig. 1(a)) and the average charge current densities defined as $\langle j_c^{Au}\rangle = (1/d_{Au}) \int_0^{d_{Au}} j_c^{Au}(z)dz$ and $\langle j_c^{Ge}\rangle = (1/d_{Ge}) \int_{d_{Au}}^{d_{Au}+d_{Ge}} j_c^{Ge}(z)dz$ and given by

$$\langle j_c^{Au}\rangle = \theta_{SH}^{Au} \left(\frac{2e}{\hbar}\right) \frac{\lambda_{Au}}{d_{Au}} \left( \frac{\frac{g_{Ge}}{g_{Au}} \tanh d_{Ge}/\lambda_{Ge} + \tanh d_{Au}/2\lambda_{Au}}{\frac{g_{Ge}}{g_{Au}} \tanh d_{Ge}/\lambda_{Ge} \coth d_{Au}/\lambda_{Au} + 1} \right) j_{s0}, \tag{11}$$

$$\langle j_c^{Ge}\rangle = \theta_{SH}^{Ge}\left(\frac{2e}{\hbar}\right)\frac{g_{Ge}}{g_{Au}}\frac{\lambda_{Ge}}{d_{Ge}}\left(\frac{\tanh d_{Ge}/2\lambda_{Ge}\tanh d_{Ge}/\lambda_{Ge}}{\frac{g_{Ge}}{g_{Au}}\tanh d_{Ge}/\lambda_{Ge}\cosh d_{Au}/\lambda_{Au}+\sinh d_{Au}/\lambda_{Au}}\right)j_{s0}. \tag{12}$$

From these equations is clear that for $g_{Ge}/g_{Au} \sim 10^{-4} \ll 1$, the spin current is almost entirely absorbed into the Au layer, and the total charge current does not depend of $\theta_{SH}^{Ge}$. Taking $\theta_{SH}^{Ge} = 0.001$ [9], the values previously obtained for $g_{eff}^{\uparrow\downarrow}(Sn = 0\%)$ and $\lambda_{Au}$, and the experimental value of $I_c = 24.8\ nA$ for $Sn = 0\%$ in Fig. 2(c) we obtain the spin Hall angle of the Au layer of $\theta_{SH}^{Au} = 0.007$ [30,31].

After having characterized the spin pumping and ISHE in the NiFe(15nm)/Au(5nm)/Ge(40nm) sample, we proceed to investigate the spin-to-charge current conversion in NiFe(15nm)/Au(5nm)/Ge$_{1-x}$Sn$_x$ samples. Specifically, we want to determine the spin Hall angle of the Ge$_{1-x}$Sn$_x$ semiconductor as a function of the Sn content. For this purpose, for each Sn concentration, we follow the procedure described above, with the exception that now, the conductivity ($\sigma_{GeSn}$) and the spin diffusion length ($\lambda_{GeSn}$) of the Ge$_{1-x}$Sn$_x$ layer are unknown parameters. In this case, we begin estimating the Ge$_{1-x}$Sn$_x$ conductivity, as we explain below. The average current densities in $Au$ and Ge$_{1-x}$Sn$_x$ are given by Eqs. (11) and (12) changing Ge by GeSn. With the values of $\sigma_{GeSn}$, and $\alpha_{Sn\%} = \alpha_0 + \alpha'$, the latter taken from the experimental data in Fig.1(d), we obtain the spin diffusion length $\lambda_{GeSn}$. Then, from the fit of the ISHE current in Fig. 2(c), the spin Hall angle $\theta_{SH}^{GeSn}$ can be extracted for each Sn concentration.

As the electrical conductivity of the Ge$_{1-x}$Sn$_x$ is presently unknown, at first glance, we can assume the same conductivity of the Ge layer, i.e., $\sigma_{GeSn} = \sigma_{Ge} = 2.7 \times 10^4\ (\Omega\ m)^{-1}$ [32]. In this case, for an Sn concentration of 6 %, with $\alpha = 7.88 \times 10^3$ we obtain a spin diffusion length of $\lambda_{GeSn} = 0.1\ nm$, four orders of magnitude less than that of Ge. On the other hand, when we calculate for this value of $\lambda_{GeSn}$, the spin Hall angle accounting for the increase of the ISHE current ($I_{CSn6\%} = 275\ nA$) in Fig. 2(c). we obtain an unphysical spin Hall angle of $\theta_{SH}^{GeSn} = 1.6$. Thus, we have to consider $\sigma_{GeSn} > \sigma_{Ge}$ [32,33].

| $S_n(\%)$ | $\alpha(10^{-3})$ | $\sigma_{GeSn} \times 10^5 (\Omega\ m)^{-1}$ | $\lambda_{GeSn}(nm)$ | $g_{GeSn} \times 10^{18}(m^{-2})$ | $I_C(nA)$ | $\theta_{SH}^{GeSn}$ |
|---|---|---|---|---|---|---|
| 3 | 7.55 | 2.7 − 27 | 7.0 − 50 | 1 − 1.4 | 170 | 0.076 − 0.029 |
| 6 | 7.88 | 2.7 − 27 | 1.1 − 10 | 6.2 − 6.9 | 275 | 0.16 − 0.016 |
| 9 | 7.75 | 2.7 − 27 | 1.5 − 16 | 4.6 − 4.3 | 225 | 0.12 − 0.013 |
| 13 | 7.55 | 2.7 − 27 | 7.0 − 50 | 1 − 1.4 | 80 | 0.026 − 0.01 |

Table I. Estimated electrical conductivity ($\sigma_{GeSn}$), measured Gilbert damping parameter ($\alpha$) and ISHE current ($I_C$). The spin diffusion length ($\lambda_{GeSn}$) and the spin Hall angle ($\theta_{SH}^{GeSn}$) were determined by the procedure explained in the text. $g_{GeSn} = \sigma_{GeSn}h/\lambda_{GeSn}e^2$ is the conductance per spin of the Au/GeSn interface.

Table I summarizes the results obtained by taking two values of $\sigma_{GeSn}$ that are at least one and two orders of magnitude higher than that of Ge, i.e., $\sigma_{GeSn} = 2.7 \times 10^5$ and $2.7 \times 10^6$ $(\Omega\, m)^{-1}$. With $\sigma_{GeSn} = 2.7 \times 10^5$ $(\Omega\, m)^{-1}$, from 3 to 13 % of Sn concentration, the spin diffusion length varies between $1.0\, nm$ to $6.9\, nm$.

The combined increase of the electrical conductivity and decrease in spin diffusion length results in the conductance per spin $g_{GeSn}$ of the Au/GeSn interface being six orders of magnitude larger than that of the Au/Ge interface. These values account for the damping enhancement due to spin relaxation in the 40 nm GeSn layer. Additionally, the strong increase of the ISHE current when compared with the NiFe(15nm)/Au(5nm)/Ge(40nm) sample leads to spin Hall angle values that vary between 0.026 and 0.16. Notably, the magnitude obtained for $\theta_{SH}^{GeSn}$ is comparable with values found in NiFe/heavy-metal interfaces [20,31,34] and almost two orders of magnitude higher than that of Ge.

### III. CONCLUSIONS

To summarize, we investigate the spin-to-charge current conversion in NiFe/Au/Ge$_{1-x}$Sn$_x$ multilayers as a function of the Sn concentration. The spin current, generated by spin pumping-FMR in NiFe layer, was injected into Au/GeSn layers and converted into a measurable charge current through ISHE. A non-monotonic dependence of the ISHE-induced current and Gilbert damping parameter on Sn concentration was observed, with a pronounced peak at 6 % Sn. Through the analysis of the magnetic damping and induced charge currents, the conductivity per spin at the Au/GeSn interface $g_{GeSn}$ and the spin Hall angle $\theta_{SH}^{GeSn}$ were estimated. The values of $\theta_{SH}^{GeSn}$ are in the range of those found in heavy metals such as Pt and Ta. We have also shown that it is possible to improve the spin conductivity at the Au/GeSn interface via variation of the Sn molar fraction. These findings establish GeSn alloys as a promising platform for integrating spintronic functionality into mainstream semiconductor technologies.

**ACKNOWLEDGMENTS**

The research at the Pontificia Universidad de Chile was supported by Fondo Nacional de Desarrollo Científico y Tecnológico (FONDECYT) Grant No. 1210641, and FONDEQUIP projects EQM180103 and EQM190136. The research at the Universidad de Santiago de Chile was supported by ANID PIA/APOYO AFB220001, project EQM230104 and 042113OM_Postdoc. C.G-F acknowledges support by FONDECYT 11220854 and PIA/APOYO AFB230003. F.P. acknowledges support by the Air Force Office of Scientific Research under the award number FA8655-22-1-7050.